\documentclass[twocolumn,aps,pra]{revtex4-2}
\usepackage{amsfonts}
\usepackage{amssymb}
\usepackage{amsmath}
\usepackage{bm}
\usepackage{color}
\usepackage{epsfig}
\usepackage{ifthen}
\usepackage{graphicx}
\usepackage{multirow}
\usepackage{bigstrut}
\usepackage{rotating}
\usepackage{longtable}
\usepackage[hidelinks,colorlinks=true,linkcolor=blue,citecolor=blue,filecolor=blue,urlcolor=blue]{hyperref}
\usepackage[utf8]{inputenc}
\usepackage[T1]{fontenc}

\def\be{ \begin{equation} }
\def\ee{ \end{equation} }
\def\bea{ \begin{eqnarray} }
\def\eea{ \end{eqnarray} }
\def\bse{ \begin{subequations} }
\def\ese{ \end{subequations} }

\def\ket#1{\,| #1 \rangle}

\def\U{\hat{U}}
\def\U{\mathbf{U}}

\def\fromto{\leftrightarrow}
\def\deg{\circ}
\def\eps{\epsilon}

\begin{document}

\author{Hayk L. Gevorgyan and Nikolay V. Vitanov}
\affiliation{Department of Physics, St. Kliment Ohridski University of Sofia, 5 James Bourchier blvd, 1164 Sofia, Bulgaria}

%\title{Composite sequences for ultrasmall transition probability: Application to deterministic single-photon emission }
\title{Deterministic generation of arbitrary ultrasmall excitation of quantum systems by composite pulse sequences}

\date{\today}

\begin{abstract}
In some applications of quantum control, it is necessary to produce very weak excitation of a quantum system.
Such an example is presented by the concept of single-photon generation in cold atomic ensembles or doped solids, e.g. by the DLCZ protocol, for which a single excitation is shared among thousands and millions atoms or ions.
Another example is the possibility to create huge Dicke state of $N$ qubits sharing a single or a few excitations.
Other examples are using tiny rotations to tune high-fidelity quantum gates or using these tiny rotations for testing high-fidelity quantum process tomography protocols. 
Ultrasmall excitation of a quantum transition can be generated by either a very weak or far-detuned driving field.
However, these two approaches are sensitive to variations in the experimental parameters, e.g. the transition probability varies with the square of the pulse area.
Here we propose a different method for generating a well-defined pre-selected very small transition probability --- of the order of $10^{-2}$ to $10^{-8}$ --- by using composite pulse sequences.
The method features high fidelity and robustness to variations in the pulse area and the pulse duration.
\end{abstract}

\maketitle
%\cite{}
%%%%%%%%%%%%%%%%%%%%%%%%%%%%%%%%%%%%%%%%%%%%%%%%%%%%%%%%%%%%%%%%%%%%%%%%%%%%%%%%%%%%%%%%%%%%%%%%%%%%%%%%%%%%%%%%%%%%%%%%%%%%%%%%%%%%%%%%%
%%%%%%%%%%%%%%%%%%%%%%%%%%%%%%%%%%%%%%%%%%%%%%%%%%%%%%%%%%%%%%%%%%%%%%%%%%%%%%%%%%%%%%%%%%%%%%%%%%%%%%%%%%%%%%%%%%%%%%%%%%%%%%%%%%%%%%%%%
%%%%%%%%%%%%%%%%%%%%%%%%%%%%%%%%%%%%%%%%%%%%%%%%%%%%%%%%%%%%%%%%%%%%%%%%%%%%%%%%%%%%%%%%%%%%%%%%%%%%%%%%%%%%%%%%%%%%%%%%%%%%%%%%%%%%%%%%%

%%%%%%%%%%%%%%%%%%%%%%%%%%%%%%%%%%%%%%%%%%%%%%%%%%%%%%%%%%%%%%%%%%%%%%%%%%%%%%%%%%%%%%%%%%%%%%%%%%%%%%%%%%%%%%%%%%%%%%%%%%%%%%%%%%%%%%%%%
\section{Introduction}
%%%%%%%%%%%%%%%%%%%%%%%%%%%%%%%%%%%%%%%%%%%%%%%%%%%%%%%%%%%%%%%%%%%%%%%%%%%%%%%%%%%%%%%%%%%%%%%%%%%%%%%%%%%%%%%%%%%%%%%%%%%%%%%%%%%%%%%%%

%Quantum control of quantum systems with two or three states is almost always focused at the generation of full or half excitation, referred to as $X$ and Hadamard (or $\sqrt{X}$) gates in quantum information, respectively.

In almost all applications of quantum control, the focus is either on complete population inversion (known as $X$ gate in quantum information) or half excitation (known as Hadamard or $\sqrt{X}$ gate in quantum information).
These are produced most often by resonant excitation by $\pi$ and $\pi/2$ pulses, but adiabatic and composite methods have also been used.
These methods have different advantages and shortcomings.
For instance, resonant excitation is the fastest method and is very accurate if the parameter values are very precise \cite{Shore1990,Shore2011}, but it is sensitive to parameter variations.
Adiabatic methods \cite{Vitanov2001,Vitanov2017} are robust to experimental errors but are slow and it is difficult to reach high accuracy with them.
(A cure is offered by the ``shortcuts-to-adiabaticity'' approach \cite{STA}, but it comes with the necessity of accurate pulse shaping or additional fields.)
Composite pulses --- trains of pulses with well-defined relative phases used as control parameters \cite{Levitt1986,Levitt2007} ---  sit somewhere in the ``sweet spot'' as they feature extreme accuracy and robustness, while being significantly faster than adiabatic methods (but slower than resonant excitation by a factor of 2-3 or more).

However, quantum control offers the opportunity for partial excitation with any transition probability, rather than just 1 and $\frac12$.
%For instance, on exact resonance, the transition probability is $P=\sin^2(A/2)$, where $A$ is the temporal pulse area --- the time integral over the Rabi frequency.
%While resonant excitation is fast, it is also sensitive to variations in the pulse area $A$ as well as the detuning from resonance $\Delta$.
%There are well established methods which can produce any desired excitation in a robust fashion, the most prominent and used being the adiabatic and composite methods.
%
For instance, there are applications in which a very small transition probability is required.
One prominent example is the DLCZ protocol for single-photon generation in an ensemble of ultracold atoms or in a doped solid and its variations and extensions \cite{DLCZ,Lvovsky2009,Sangouard2011,Yang2016,Pu2017,Laplane2017,Kutluer2017}.
Single photons are the physical platform for such advanced technologies as quantum communications \cite{Gisin2007,Ursin2007,Kimble2008,Zhang2017,Chen2021} and photonic quantum computing \cite{Kok2007,Barz2015,Rudolph2017,Takeda2019}. 
In this protocol, a three-level Raman system $\ket{g_1} \fromto \ket{e} \fromto \ket{g_2}$  is used.
In the \emph{writing} process, the atomic transition $\ket{g_1} \fromto \ket{e}$ is excited with a very low probability by an off-resonant laser pulse with a wave vector $\overrightarrow{k_w}$, such that a single (or a few) atomic excitation is stored in the ensemble as a shared excitation by all atoms.
Then collective spontaneous emission on the transition  $\ket{e} \to \ket{g_2}$ occurs at a random time, in which a (Stokes) photon is emitted in a random direction.
However, a single-photon detector is placed along a particular spatial direction and any click in it is considered as a ``heralded'' photon, with a well-defined wave vector $\overrightarrow{k_s}$.
In the \emph{reading} process, a resonant laser pulse with a wave vector $\overrightarrow{k_r}$ is applied on the atomic transition $\ket{g_2} \fromto \ket{e}$, which stimulates the emission of a (anti-Stokes) photon on the pump transition $\ket{e} \to \ket{g_1}$, in a well-defined spatial direction $\overrightarrow{k_a}$, determined by the phase-matching condition
%\be\label{phase matching}
$\overrightarrow{k_a} = \overrightarrow{k_s} + \overrightarrow{k_w} - \overrightarrow{k_r}$.
%\ee
In this protocol, one of the crucial conditions is to be able to produce only one shared excitation among a large number of atoms $N$, i.e. a driving field which generates a transition probability of $1/N$ is needed.

Another example is the possibility to create huge entangled Dicke states \cite{Dicke1954}.
These very special states share a fixed number of excitations $n$ evenly among $N$ qubits, a special case of which (for $n=1$) is the W state. 
A prominent feature of the Dicke states is that they are immune against collective dephasing, which is ubiquitous in various systems. 
Therefore, the Dicke sub-space, which is $N!/n!(N-n)!$-dimensional, can be used as a decoherence-free computational subspace \cite{Ivanov2008,Ivanov2010,Linington2009}. 
%Dicke states generalize W states, which can be used for quantum communication [7]. 
Dicke states possess genuine multi-partite entanglement \cite{Toth2007,Usha2007}, which is, moreover, very robust against particle loss \cite{Stockton2003,Bourennane2006,Dur2001}: the loss of a qubit reduces the $N$-dimensional Dicke state to a $N-1$-dimensional one. 
%Due to their robust entanglement, these states are particularly well suited for the experimental examination of multi-partite entanglement and can be used to test fundamental concepts of quantum mechanics.
Dicke states have been proposed and demonstrated in various physical systems, including ensembles of neutral atoms \cite{Stockton2004, Thiel2007}, trapped ions \cite{Haffner2005,Hume2009,Linington2008,Ivanov2013}, quantum dots \cite{Zou2003},
and using linear optics \cite{Kiesel2007,Thiel2007}. 
Many of these proposals and demonstrations have various restrictions, as they cannot create arbitrary but only
particular Dicke states, individual qubit addressing is required, the number of the necessary physical interactions scales very fast with $N$, a special initial (Fock) state is required, insufficient efficiency, very long interaction times, etc.
Composite pulses of ultrasmall probability offer a direct path toward the creation of large Dicke states as they can produce a specific number of shared excitations among large-$N$ ensembles of qubits.

A third example when a well-defined small transition probability is needed arises when fine tuning quantum gates: in order to reach ultrahigh gate fidelity a rotation gate at a well-defined tiny angle can be very useful. Moreover, such small rotations alone can be used to test the accuracy of various quantum process tomography protocols.

In this paper, we address this specific problem by designing composite pulse sequences, which seem to be the only quantum control technique capable to generate a tiny transition probability that is robust to variations of the experimental parameters. 
The dominant majority of composite pulses in the literature are designed to produce specific rotations on the Bloch sphere, typically at angles $\pi$ (generating complete population transfer), $\pi/2$ (half population transfer), $\pi/4$ and $3\pi/4$, as reviewed in Refs.~\cite{Levitt1986,  Levitt2007}.
There exist just a few composite sequences which produce general rotations at arbitrary angles \cite{Wimperis1990, Wimperis1991, Wimperis1994, Torosov2019, Torosov2020, Gevorgyan2021}. Some of them can be used for the present task of ultrasmall probability and they are listed below, along with many newly derived composite sequences.

Composite rotations are broadly divided into two large groups called variable and constant rotations. 
The variable rotations \cite{Levitt1986, Torosov2019, Torosov2020} feature well-defined transition probability but not well-defined phases of the propagator. Constant (or phase-distortionless rotations) feature both well-defined populations and well-defined phases of the propagator \cite{Wimperis1990,Wimperis1991,Wimperis1994}. 
There are large markets for either of these, with only constant rotations being suitable for quantum gates.
However, they are much more demanding to generate and much longer than variable rotations, for the same order of error compensation.
This will be clearly seen below as we consider one type of constant rotations and two types of variable rotations.

After a description of the derivation method we present specific composite sequences of 2, 3 and 4 pulses, many of which have analytic expressions for the composite parameters, and then proceed to longer sequences.

%%%%%%%%%%%%%%%%%%%%%%%%%%%%%%%%%%%%%%%%%%%%%%%%%%%%%%%%%%%%%%%%%%%%%%%%%%%%%%%%%%%%%%%%%%%%%%%%%%%%%%%%%%%%%%%%%%%%%%%%%%%%%%%%%%%%%%%%%
\section{The method}
%%%%%%%%%%%%%%%%%%%%%%%%%%%%%%%%%%%%%%%%%%%%%%%%%%%%%%%%%%%%%%%%%%%%%%%%%%%%%%%%%%%%%%%%%%%%%%%%%%%%%%%%%%%%%%%%%%%%%%%%%%%%%%%%%%%%%%%%%

We wish to construct composite pulses, which produce a very low probability of transition between two states $\ket{1} \to \ket{2}$, in an efficient and robust manner.
Such composite pulses are known as $\theta$-pulses, as they produce a transition probability $p = \sin^2(\theta/2)$.
In the NMR literature one can find a number of $\theta$ pulses for $\theta = \pi/4$ (called $45^\deg$ pulses), $\theta = \pi/2$ (called $90^\deg$ pulses), and  $\theta = 3\pi/4$ (called $135^\deg$ pulses).
Very few general formulae for an arbitrary value of $\theta$ exist in the literature.
%, except for the three-pulse \cite{Wimperis1990} and five-pulse BB1 sequence \cite{Wimperis1994} of Wimperis.
%
In our case we need composite pulses, which produce transition probability $p = 1/N \ll 1$, which implies $\theta \ll 1$.
Such composite pulses are designed here.

%Because it is the value of $p$, which is important, and the phases of the created superposition are of no significance, the composite pulses shown below (except one) do not stabilize the phases of the created superposition state, i.e. these are not rotation gates in the sense of quantum information.

%The composite pulses below are robust to pulse area errors, in order to account for the spatial variation of the laser fields.
%They are not robust to detuning errors.
%Therefore, they are more suitable for the rubidium-MOT experiment, rather than the doped solids.
%Composite pulses, which are robust to both pulse area and detuning errors can be derived too (with a further effort).

Each pulse in a composite sequence is considered resonant and hence it generates the propagator
\be\label{U1}
\U (A,\phi) = \left[ \begin{array}{cc} \cos(A/2) & -i e^{i\phi} \sin(A/2) \\ -i e^{-i\phi} \sin(A/2) & \cos(A/2)  \end{array}  \right],
\ee
where $\phi$ is the phase of the coupling.
The overall propagator for a sequence of $n$ pulses,
\be\label{sequence}
(A_1)_{\phi_1} (A_2)_{\phi_2} \cdots (A_n)_{\phi_n},
\ee
each with a pulse area $A_k$ and phase $\phi_k$, reads
\be\label{Un}
\U_n = \U (A_n,\phi_n) \U (A_{n-1},\phi_{n-1}) \cdots \U (A_2,\phi_2) \U (A_1,\phi_1),
\ee
which, by convention, acts from right to left.
One of the phases is always irrelevant for the physically observed quantities (it is related to the global phase of the wavefunction), and can be set to zero.
As such, we always choose the first phase: $\phi_1 = 0$.
In other words, all other phases are relative phases of the respective pulse to the phase of the first pulse.

The pulse areas $A_k$ and the phases $\phi_k$ are the control parameters, which are selected from the conditions that the transition probability,
\be\label{p}
P = |\U_{12}|^2,
\ee
has a specific target value $p$ and it is robust to variations $\eps$ in the pulse area $A_k (1+\eps)$.
The error-free values of the pulse areas $A_k$ are called \emph{nominal} values.
The relative error $\eps$ is assumed to be the same for all pulses in the composite sequence.
This is reasonable if they are derived from the same source, which is usually the case.

The multiplication of the two-dimensional matrices in Eq.~\eqref{Un} leads to rapidly growing expressions.
Still, these are far more manageable than the ones coming from the three-dimensional matrices in the usual Bloch-vector derivation of composite sequences.

One can proceed in two directions.
\begin{itemize}

\item
One possibility is to expand the transition probability of Eq.~\eqref{p} in a Taylor-Maclaurin series vs $\eps$.
The coefficients in this series are functions of all $A_k$ and $\phi_k$ ($k = 1, 2, \ldots, n$).
We nullify as many of the first few such coefficients (i.e. derivatives vs $\epsilon$) as possible, which generate a set of equations for $A_k$ and $\phi_k$.
The result is a transition probability with a Taylor-Maclaurin series expansion
\be\label{p-series}
P(\eps) = p + O(\eps^m),
\ee
where $p$ is the target value.
We say that the respective composite sequence is accurate up to order $O(\eps^m)$.
We shall first present such composite sequences, which are known as \textit{variable rotations} in NMR and allow to easily reach error compensation of very high order.

\item
Alternatively, one can take the propagator elements $U_{11} = U_{22}^*$ and $U_{12} = -U_{21}^*$, expand them in Taylor-Maclaurin series vs $\eps$, and carry out elimination of as many lowest-order terms as possible.
The result is a Taylor-Maclaurin expansion of the propagator,
\be\label{U-series}
\U_n(\eps) = \U_n +  O(\eps^l).
\ee
Obviously, with the same number of free parameters, one can cancel of factor of 2 fewer terms now, than in the expansion of the probability $P$.
However, the resulting composite sequences will be stabilized with respect to both the amplitudes and the phases of the overall propagator, rather than with respect to the amplitudes only, as with Eq.~\eqref{p-series}.
Such composite sequences create constant rotations in NMR language, or, in quantum information terms, \emph{quantum rotation gates}.
%The five-pulse composite sequence BB1 of Wimperis (1990) is a rotation gate that is accurate up to order $O(\eps^3)$ in $\U$.

\end{itemize}

We begin with the first approach, which delivers expressions as in Eq.~\eqref{p-series}, and then proceed with the second approach, which delivers expressions of the type \eqref{U-series}.

%%%%%%%%%%%%%%%%%%%%%%%%%%%%%%%%%%%%%%%%%%%%%%%%%%%%%%%%%%%%%%%%%%%%%%%%%%%%%%%%%%%%%%%%%%%%%%%%%%%%%%%%%%%%%%%%%%%%%%%%%%%%%%%%%%%%%%%%%
\section{Small-probability composite sequences}
%%%%%%%%%%%%%%%%%%%%%%%%%%%%%%%%%%%%%%%%%%%%%%%%%%%%%%%%%%%%%%%%%%%%%%%%%%%%%%%%%%%%%%%%%%%%%%%%%%%%%%%%%%%%%%%%%%%%%%%%%%%%%%%%%%%%%%%%%

%%%%%%%%%%%%%%%%%%%%%%%%%%%%%%%%%%%%%%%%%%%%%%%%%%%%%%%%%%%%%%%%%%%%%%%%%%%%%%%%%%%%%%%%%%%%%%%%%%%%%%%%%%%%%%%%%%%%%%%%%%%%%%%%%%%%%%%%%
\subsection{Two-pulse composite sequences}
%%%%%%%%%%%%%%%%%%%%%%%%%%%%%%%%%%%%%%%%%%%%%%%%%%%%%%%%%%%%%%%%%%%%%%%%%%%%%%%%%%%%%%%%%%%%%%%%%%%%%%%%%%%%%%%%%%%%%%%%%%%%%%%%%%%%%%%%%

We have derived two types of two-pulse composite sequences.

%%%%%%%%%%%%%%%%%%%%%%%%%%%%%%%%%%%%%%%%%%%%%%%%%%%%%%%%%%%%
\subsubsection{Symmetric sequence of pulses}

In the first type, the two pulse areas are equal to $\pi/2$,
\be\label{CP2a}
%\alpha_0 \alpha_\phi,
S2:\quad (\tfrac12\pi)_0 (\tfrac12\pi)_{\pi-\theta}.
\ee
%with $\eta = \text{arccos} (2p-1)$.
The transition probability is
\be
P = \cos^2 \frac{\pi\eps}{2} \sin^2 \frac{\theta}{2}.
\ee
For 
\be
\theta = \text{arccos} (1-2p) = 2\arcsin(\sqrt{p}),
\ee
we find
\be
P = p [1-\sin^2 (\frac12\pi\eps)] = p [1 + O(\eps^2)].
\ee
This simplest composite sequence is accurate up to the second order $O(\eps^2)$.
For example, for probabilities $p = 10^{-2}$, $10^{-3}$, $10^{-4}$ and $10^{-5}$ we find $\phi = 0.0638\pi$, $0.0201\pi$, $0.0064\pi$, and $0.0020\pi$.
These values correspond to $11.48^\deg$, $3.62^\deg$, $1.15^\deg$, and $0.36^\deg$.

%An error in the phase, $\phi \to \phi + \delta$, gives
%\be
%P \approx p \cos^2 \frac{\pi\eps}{2} \left[1 + \delta\tan\phi + O(\delta^2) \right].
%\ee
%Because the values of $\phi$ are close to $0$, we have $|\tan\phi| \ll 1$, and hence this composite sequence is not terribly sensitive to phase errors.
%For example, errors $\delta$ of $0.1^\deg$ and $0.01^\deg$  in the phase $\phi$ for the probability of $10^{-4}$ lead to relative probability errors of $5.5\%$ and $0.6\%$.

The advantage of these sequences is their extreme simplicity and the analytic formula for the phase, which make it possible to immediately write down the sequence for any target transition probability. 
The disadvantage is the availability of a single control parameter only, which limits the error compensation to the first order only. 
This is still superior over a single resonant pulse, which is accurate to zeroth order only.

%%%%%%%%%%%%%%%%%%%%%%%%%%%%%%%%%%%%%%%%%%%%%%%%%%%%%%%%%%%%
\subsubsection{Asymmetric sequence of pulses}

In the second two-pulse sequence, the pulse areas are different,
\be\label{CP2b}
A2:\quad (A_1)_0 (A_2)_{\phi_2}.
\ee
Here we have three control parameters --- two pulse areas and a phase --- which allow us to compensate higher orders of errors. 
Now closed analytic expressions for the parameters are not possible to derive.
However, due to the fact that $p \ll 1$, we can use perturbation theory, which gives us the approximations
\be
 A_1 = x - y,\quad A_2 = x + y,\quad \phi_2 = \pi - \phi,
\ee
with
 $x \approx 0.7151\pi$, $y \approx 0.2553 \pi \sqrt{p}$, and $\phi \approx 0.4875 \pi \sqrt{p}$.
All these are valid for $p \ll 1$.
The pulse areas and the phases for a few values of the transition probability are given in Table \ref{Table:CP2}.

The advantage of the composite sequence \eqref{CP2b} over the symmetric one \eqref{CP2a} is that it is accurate to the third order in $\eps$,
\be
P = p [1 + O(\eps^3)].
\ee
The disadvantage is that it requires a larger total pulse area, about $1.43\pi$ compared to just $\pi$ for the symmetric sequence \eqref{CP2a}.

%T%T%T%T%T%T%T%T%T%T%T%T%T%T%T%T%T%T%T%T%T%T%T%T%T%T%T%T%T%T%T%T%T%T%T%T%T%T%T%T%T%T%T%T%T%T%T%T
\begin{table}[tb]
\centering
\begin{tabular}{|l|cc|c|}
\hline
$p$ & $A_1$ & $A_2$ & $\phi$ \\ \hline
%\multicolumn{5}{|c|}{Symmetric sequences} \\ \hline 
%$10^{-1}$ & 2 & 0.5 & 0.5 & 0.795167 \\
%$10^{-2}$ & 2 & 0.5 & 0.5 & 0.063769 \\
%$10^{-3}$ & 2 & 0.5 & 0.5 & 0.020135 \\
%$10^{-4}$ & 2 & 0.5 & 0.5 & 0.006366 \\
%$10^{-5}$ & 2 & 0.5 & 0.5 & 0.002013 \\
%$10^{-6}$ & 2 & 0.5 & 0.5 & $6.37 \times 10^{-4}$ \\
%$10^{-7}$ & 2 & 0.5 & 0.5 & $2.01 \times 10^{-4}$ \\
%$10^{-8}$ & 2 & 0.5 & 0.5 & $6.37 \times 10^{-5}$ \\
%\hline
%\multicolumn{5}{|c|}{Asymmetric sequences} \\ \hline 
%$10^{-1}$ & 3 & 0.633668 & 0.803479 & 0.845510 \\
$10^{-2}$ &  0.689806 & 0.741105 & 0.048767 \\
$10^{-3}$ &  0.707103 & 0.723255 & 0.015417 \\
$10^{-4}$ &  0.712599 & 0.717704 & 0.004875 \\
$10^{-5}$ &  0.714341 & 0.715956 & 0.001542 \\
$10^{-6}$ &  0.714894 & 0.715404 & $4.88 \times 10^{-4}$ \\
$10^{-7}$ &  0.715068 & 0.715229 & $1.54 \times 10^{-4}$ \\
$10^{-8}$ &  0.715123 & 0.715174 & $4.88 \times 10^{-5}$ \\
%$10^{-9}$ & 0.715141 & 0.715157 & 0.999980 \\
%$10^{-10}$ & 0.715145 & 0.715150 & 1 \\
\hline
\end{tabular}
\caption{Pulse areas and phases (in units of $\pi$) for the composite sequence \eqref{CP2b} (in units of $\pi$) for a few values of the transition probability. 
All composite sequences have the error order $O(\epsilon^3)$.
%Also given is the order $n$ of pulse area error compensation $O(\epsilon^n)$, indicating that all orders below $n$ are zero.
 }
\label{Table:CP2}
\end{table}
%T%T%T%T%T%T%T%T%T%T%T%T%T%T%T%T%T%T%T%T%T%T%T%T%T%T%T%T%T%T%T%T%T%T%T%T%T%T%T%T%T%T%T%T%T%T%T%T

\begin{figure}[tb]
\begin{tabular}{r}
    \includegraphics[width=0.90\columnwidth]{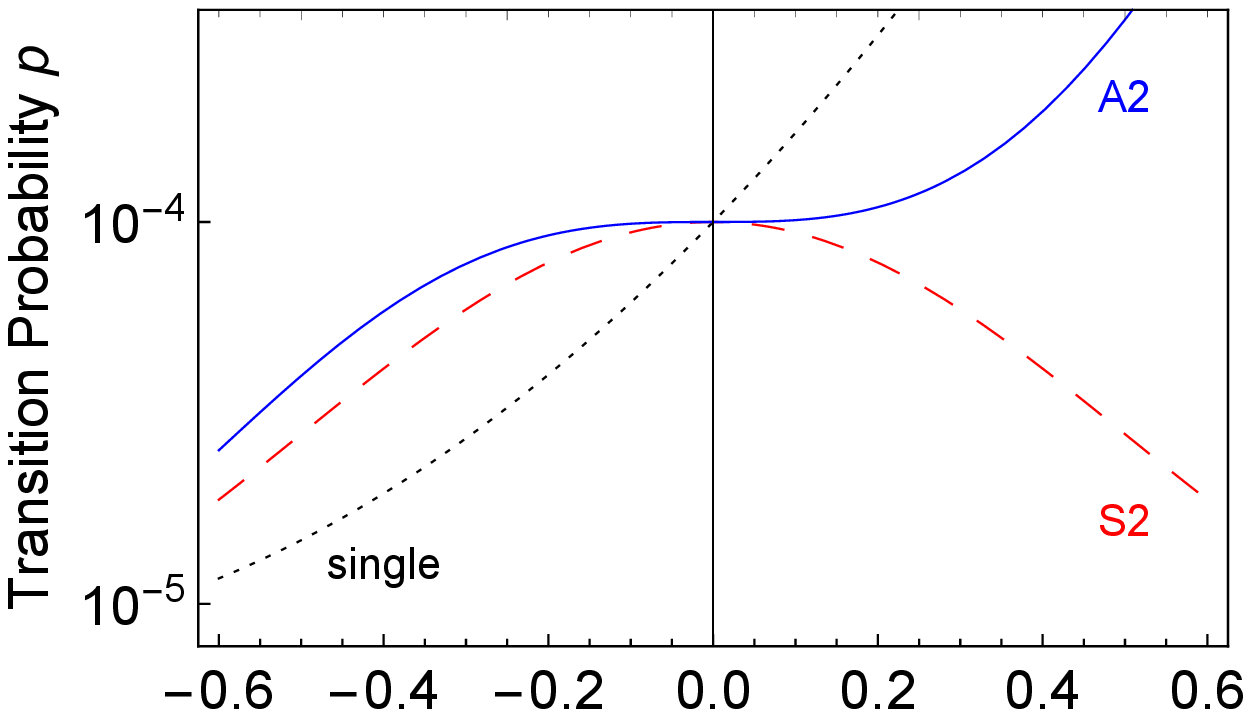} \\
    \includegraphics[width=0.90\columnwidth]{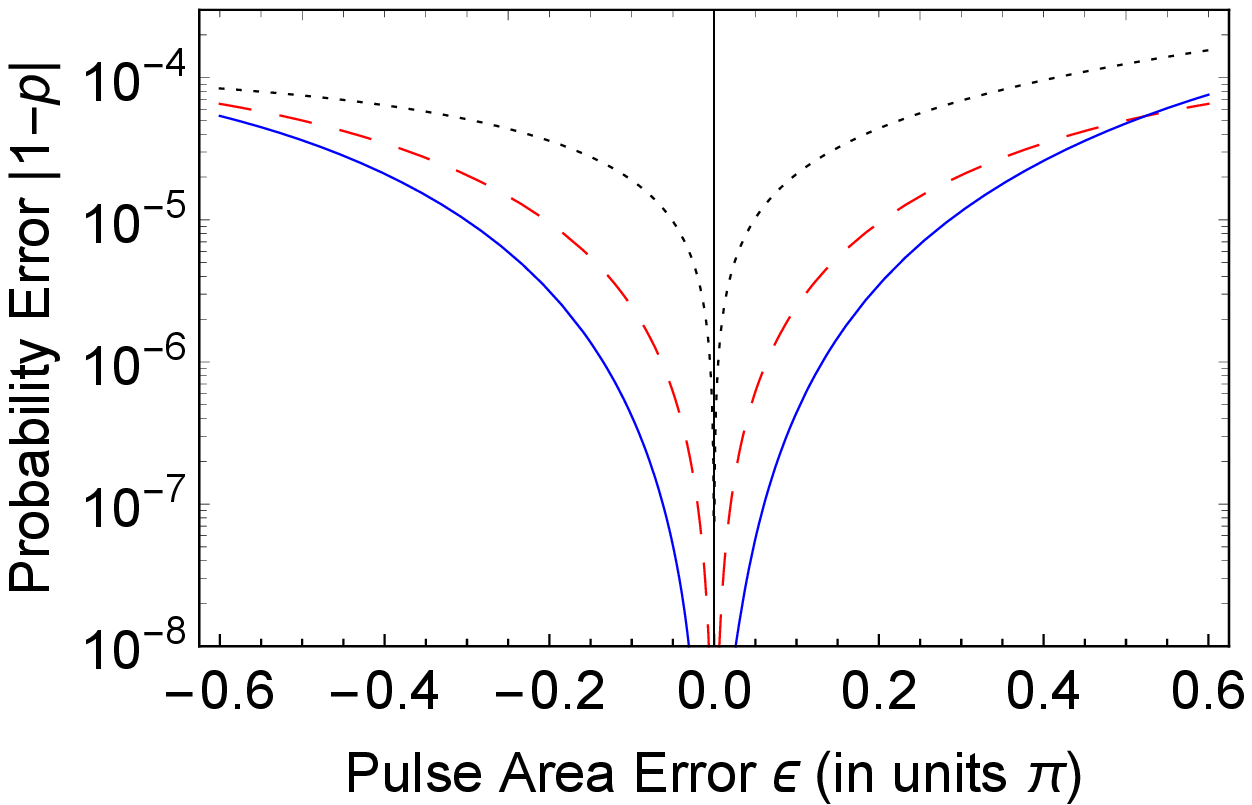}
\end{tabular}
\caption{
Performance of the two-pulse composite sequences \eqref{CP2a} (red dashed) and \eqref{CP2b} (blue solid) for the transition probability $p=10^{-4}$.
The dotted curves show the single pulse excitation probability for comparison.
}
\label{Fig:cp2}
\end{figure}

The performance of the two sequences is compared in Fig.~\ref{Fig:cp2}.
Both sequences \eqref{CP2a} and \eqref{CP2b} outperform significantly the conventional single-pulse excitation probability, which is very sensitive to pulse area errors.
The asymmetric sequence A2 of Eq.~\eqref{CP2b}, with its three control parameters and error order $O(\epsilon^3)$, outperforms the symmetric sequence S2 of Eq.~\eqref{CP2a}, which has only one control parameter and error order $O(\epsilon^2)$.

%F%F%F%F%F%F%F%F%F%F%F%F%F%F%F%F%F%F%F%F%F%F%F%F%F%F%F%F%F%F%F%F%F%F%F%F%F%F%F%F%F%F%F%F%F%F%F%F%F%F%F%F%F%F%F%F
%F%F%F%F%F%F%F%F%F%F%F%F%F%F%F%F%F%F%F%F%F%F%F%F%F%F%F%F%F%F%F%F%F%F%F%F%F%F%F%F%F%F%F%F%F%F%F%F%F%F%F%F%F%F%F%F

%%%%%%%%%%%%%%%%%%%%%%%%%%%%%%%%%%%%%%%%%%%%%%%%%%%%%%%%%%%%%%%%%%%%%%%%%%%%%%%%%%%%%%%%%%%%%%%%%%%%%%%%%%%%%%%%%%%%%%%%%%%%%%%%%%%%%%%%%
\subsection{Three-pulse composite sequences}
%%%%%%%%%%%%%%%%%%%%%%%%%%%%%%%%%%%%%%%%%%%%%%%%%%%%%%%%%%%%%%%%%%%%%%%%%%%%%%%%%%%%%%%%%%%%%%%%%%%%%%%%%%%%%%%%%%%%%%%%%%%%%%%%%%%%%%%%%

We have derived three three-pulse composite sequences, two symmetric and one asymmetric.

%%%%%%%%%%%%%%%%%%%%%%%%%%%%%%%%%%%%%%%%%%%%%%%%%%%%%%%%%%%%
\subsubsection{Symmetric sequence of pulses}

The symmetric sequence of pulses reads
\be\label{CP3a}
%\alpha_0 \beta_{\phi} \alpha_{\chi},
%S3:\quad (\tfrac12{\pi})_0 \pi_{\chi} (\tfrac12{\pi})_{\psi},
S3:\quad (\tfrac12{\pi})_0 \pi_{\alpha+\beta} (\tfrac12{\pi})_{2\beta},
\ee
where 
\bse
\begin{align}
\alpha &= \theta/2 , \\
\beta &= \arccos ({\sin\alpha} - {\cos\alpha}), \\
\theta &= \arccos (1-2p) = 2\arcsin(\sqrt{p}).
\end{align}
\ese
The transition probability reads
\be
P = [1 - \sin^4(\epsilon/2)] \sin^2(\theta/2).
\ee
It is obviously accurate up to order $O(\eps^4)$. 

The sequence \eqref{CP3a} is derived as follows. 
First, we calculate the overall propagator of Eq.~\eqref{Un} for $N=3$ pulses. 
Numerical evidence suggests that the pulse areas could be taken as in Eq.~\eqref{CP3a}, i.e. a $\pi$ pulse in the middle sandwiched by two half-$\pi$ pulses.
We take the first phase to be 0, and we are left with two phases to be determined. 
The overall three-pulse transition probability for zero error ($\epsilon=0$) is readily calculated to be
\be
P = |U_{21}|^2 = \sin^2(\phi_2-\phi_3/2).
\ee
If we set $P=\sin^2 (\theta/2)$ (as for a resonant $\theta$ pulse), we find $\phi_3 = 2\phi_2 - \theta$.
Next we calculate the first few derivatives of $U_{21}$ with respect to the error $\epsilon$ and find
\begin{align}
U_{21}' (\epsilon=0) &= 0 ,\\
U_{21}'' (\epsilon=0) &= [ 1+2 \cos (\theta) + 2 \cos (\phi_2) \notag\\
&+2 \cos (\theta - \phi_2) +\cos (\theta - 2\phi_2) ] / 8 , \\
U_{21}''' (\epsilon=0) &= 0 .
\end{align}
The vanishing of the odd-order derivatives follows from the choice of symmetric pulse areas in Eq.~\eqref{CP3a}.
By setting $\phi_2=\theta/2+\beta$ the equation for $U_{21}'' (\epsilon=0)$ reduces to
\be
2 \cos \beta \cos (\theta /2) + \cos^2 \beta + \cos \theta = 0.
\ee
has 4 solutions, two complex and two real, of which one positive and one negative. 
The real positive solution is given by the expression listed in Eq.~\eqref{CP3a}.
The first nonzero derivative is $U_{21}^{(4)} (\epsilon=0)$.
%
%Table \ref{Table:CP3} presents the explicit phases for the composite sequences \eqref{CP3s} for a few values of the transition probability.
The availability of analytic formulae for the phases allows us to find their values for any value of the transition probability.

%F%F%F%F%F%F%F%F%F%F%F%F%F%F%F%F%F%F%F%F%F%F%F%F%F%F%F%F%F%F%F%F%F%F%F%F%F%F%F%F%F%F%F%F%F%F%F%F%F%F%F%F%F%F%F%F
\begin{figure}[tb]
\begin{tabular}{r}
    \includegraphics[width=0.90\columnwidth]{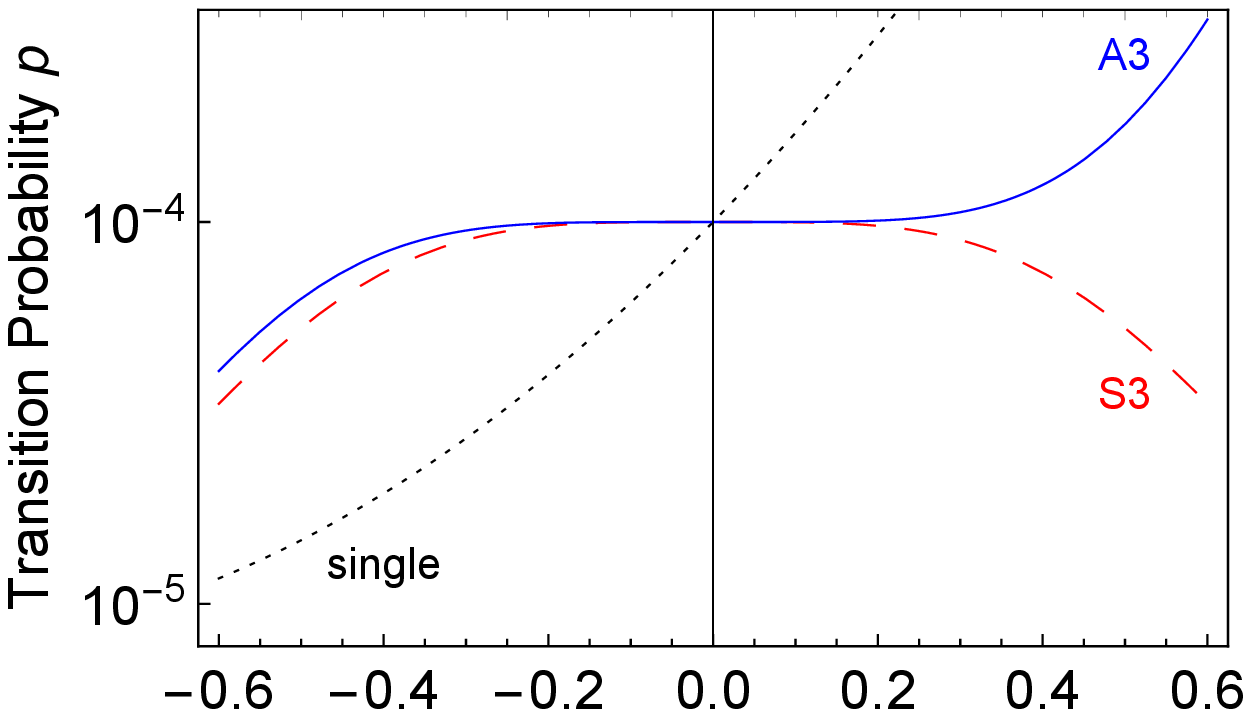} \\
    \includegraphics[width=0.90\columnwidth]{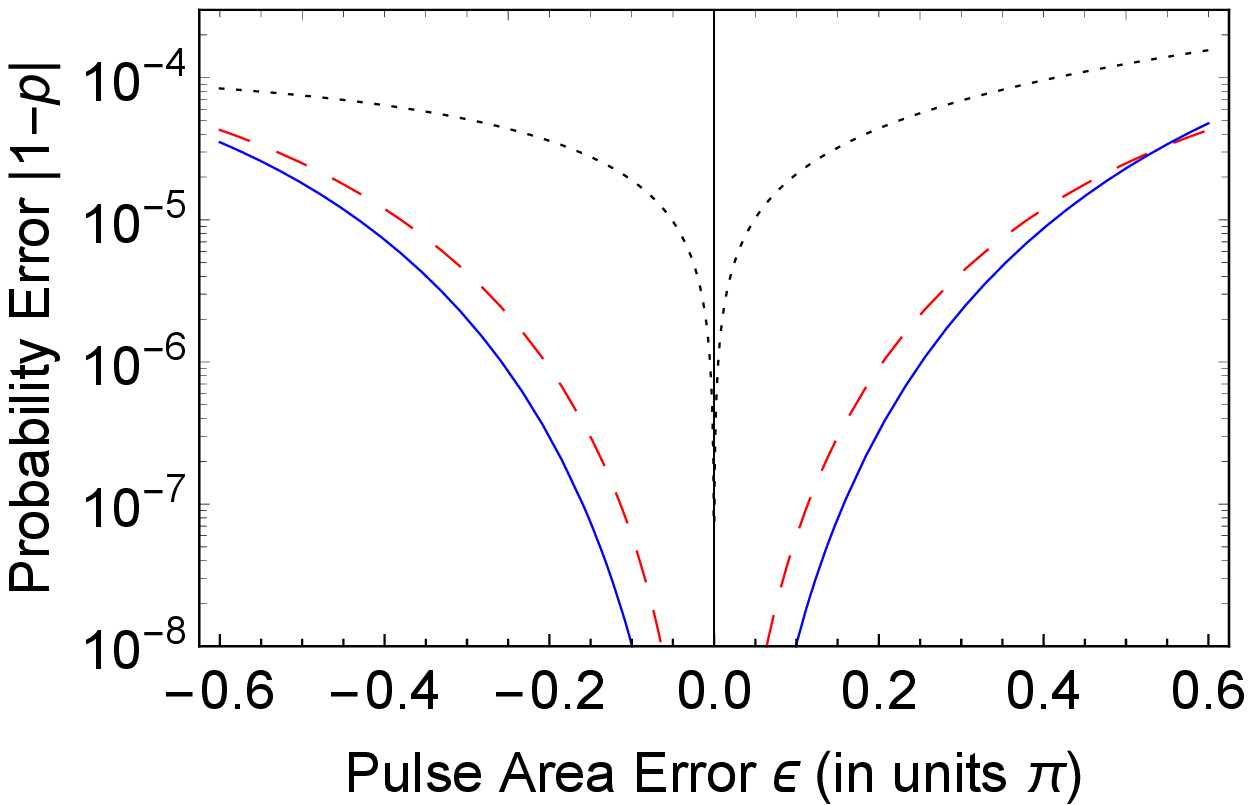}
\end{tabular}
\caption{
Performance of the three-pulse composite sequences \eqref{CP3a} (red dashed) and \eqref{CP3b} (blue solid) for the transition probability $p=10^{-4}$.
The dotted curves show the single pulse excitation probability for comparison.
}
\label{Fig:cp3}
\end{figure}
%F%F%F%F%F%F%F%F%F%F%F%F%F%F%F%F%F%F%F%F%F%F%F%F%F%F%F%F%F%F%F%F%F%F%F%F%F%F%F%F%F%F%F%F%F%F%F%F%F%F%F%F%F%F%F%F

%T%T%T%T%T%T%T%T%T%T%T%T%T%T%T%T%T%T%T%T%T%T%T%T%T%T%T%T%T%T%T%T%T%T%T%T%T%T%T%T%T%T%T%T%T%T%T%T
\begin{table}[t!]
\centering
\begin{tabular}{|l|ccc|cc|}
\hline
$p$ & $A_1$ & $A_2$ & $A_3$ & $\phi_2$ & $\phi_3$ \\
%\hline
%\multicolumn{7}{|c|}{Symmetric sequences of type S3} \\\hline 
%$10^{-1}$ & 4 & 0.5 & 1 & 0.5 & 0.8204 & 1.4359 \\
%$10^{-2}$ & 4 & 0.5 & 1 & 0.5 & 0.8847 & 1.7056 \\
%$10^{-3}$ & 4 & 0.5 & 1 & 0.5 & 0.9292 & 1.8382 \\
%$10^{-4}$ & 4 & 0.5 & 1 & 0.5 & 0.9580 & 1.9097 \\
%$10^{-5}$ & 4 & 0.5 & 1 & 0.5 & 0.9757 & 1.9493 \\
%$10^{-6}$ & 4 & 0.5 & 1 & 0.5 & 0.9861 & 1.9715 \\
%$10^{-7}$ & 4 & 0.5 & 1 & 0.5 & 0.9921 & 1.9840 \\
%$10^{-8}$ & 4 & 0.5 & 1 & 0.5 & 0.9955 & 1.9910 \\
%$10^{-9}$ & 0 & 1 & 0 & 0.997479 & 1.994937 \\
%$10^{-10}$ & 0 & 1 & 0 & 0.998580 & 1.997153 \\
%\hline
%\multicolumn{7}{|c|}{Asymmetric sequences of type W3} \\ \hline 
%$10^{-1}$ & 3 & 0.2048 & 1 & 1 & 0.510379 & 1.531136 \\
%$10^{-2}$ & 3 & 0.0638 & 1 & 1 & 0.503231 & 1.509692 \\
%$10^{-3}$ & 3 & 0.0201 & 1 & 1 & 0.501020 & 1.503060 \\
%$10^{-4}$ & 3 & 0.0064 & 1 & 1 & 0.500323 & 1.500968 \\
%$10^{-5}$ & 3 & 0.0020 & 1 & 1 & 0.500102 & 1.500306 \\
%$10^{-6}$ & 3 & $6.38 \times 10^{-4}$ & 1 & 1 & 0.500032 & 1.500097 \\
%$10^{-7}$ & 3 & $2.01 \times 10^{-4}$ & 1 & 1 & 0.500010 & 1.500031 \\
%$10^{-8}$ & 3 & $6.38 \times 10^{-5}$ & 1 & 1 & 0.500003 & 1.500010 \\
\hline
%\multicolumn{7}{|c|}{Asymmetric sequences of type A3} \\ \hline 
%$10^{-1}$ &  0.5042 & 1.3086 & 0.6321 & 1.318452 & 0.4777 \\
$10^{-2}$  & 0.5682 & 1.2436 & 0.6292 & 1.1533 & 0.2546 \\
$10^{-3}$  & 0.5904 & 1.2276 & 0.6232 & 1.0785 & 0.1405 \\
$10^{-4}$  & 0.6001 & 1.2229 & 0.6184 & 1.0419 & 0.0785 \\
$10^{-5}$  & 0.6049 & 1.2214 & 0.6151 & 1.0229 & 0.0441 \\
$10^{-6}$  & 0.6074 & 1.2209 & 0.6131 & 1.0126 & 0.0248 \\
$10^{-7}$  & 0.6087 & 1.2208 & 0.6119 & 1.0070 & 0.0139 \\
$10^{-8}$  & 0.6094 & 1.2207 & 0.6113 & 1.0039 & 0.0078 \\
\hline
\end{tabular}
\caption{Pulse areas and phases (in units of $\pi$) for the composite sequences of 3 pulses \eqref{CP3b}  for a few values of the transition probability $p$. 
%Also given is the order $n$ of pulse area error compensation $O(\epsilon^n)$, indicating that all orders below $n$ are zero.
All composite sequences have the error order $O(\epsilon^5)$.
 }
\label{Table:CP3}
\end{table}
%T%T%T%T%T%T%T%T%T%T%T%T%T%T%T%T%T%T%T%T%T%T%T%T%T%T%T%T%T%T%T%T%T%T%T%T%T%T%T%T%T%T%T%T%T%T%T%T

%%%%%%%%%%%%%%%%%%%%%%%%%%%%%%%%%%%%%%%%%%%%%%%%%%%%%%%%%%%%
\subsubsection{Asymmetric sequence of pulses}

The most general three-pulse composite sequence has the form
\be\label{CP3b}
%\alpha_0 \beta_\zeta \gamma_\upsilon.
A3:\quad (A_1)_{0} (A_2)_{\phi_2} (A_3)_{\phi_3}.
\ee
Although the composite sequence \eqref{CP3b} costs more total pulse area ($\approx 2.44\pi$) than the preceding two, it is accurate to order $O(\eps^5)$. The pulse areas and the phases computed numerically are given in Table \ref{Table:CP3}.

The performance of the three-pulse sequences is illustrated in Fig.~\ref{Fig:cp3}. 
Both sequences \eqref{CP3a} and \eqref{CP3b} outperform both the conventional single-pulse excitation probability and the two-pulse composite sequences \eqref{CP2a} and \eqref{CP2b} of Fig.~\ref{Fig:cp2}.
Moreover, the asymmetric sequence A3 of Eq.~\eqref{CP2b}, which is of error order $O(\epsilon^5)$, clearly outperforms the symmetric sequence S3 of Eq.~\eqref{CP3a}, which is of error order $O(\epsilon^4)$.

Because the three-pulse sequences seem to be the ``sweet spot'' in terms of performance (error order and high-fidelity window width) versus cost (total pulse area and control complexity), they deserve some discussion.
There are clear advantages and disadvantages of each of these two sequences. 
%The W3 sequence is phase-distortionless, i.e. constant rotation, and it is suitable for a quantum gate, or a correction to quantum gates. However, it has the lowest order of error compensation.
The S3 sequence has a nice analytic form and a total pulse area of $2\pi$. 
However, it has lower error order than A3.
The real advantage of the sequence S3 is its analytic form, which makes it very easy to calculate the composite phases for any target transition probability $p$.
The A3 sequence looks less attractive as neither the pulse area nor the phases are rational numbers and they are all numerical, but this sequence has the higher order of error compensation, although at the expense of the larger pulse area of about $2.44\pi$. 
Its real inconvenience is in the fact that for target transition probabilities not listed in Table \ref{Table:CP3} one has to calculate them numerically, although this is not a very difficult task.

%To wrap up, the recommendation is: (i) if a gate sequence is needed, use W3; (ii) if the target transition probability is different from the values in Table \ref{Table:CP3}, use S3; (iii) if the most efficient error compensation is needed use A3.

%%%%%%%%%%%%%%%%%%%%%%%%%%%%%%%%%%%%%%%%%%%%%%%%%%%%%%%%%%%%%%%%%%%%%%%%%%%%%%%%%%%%%%%%%%%%%%%%%%%%%%%%%%%%%%%%%%%%%%%%%%%%%%%%%%%%%%%%%
\subsection{Four-pulse composite sequences}
%%%%%%%%%%%%%%%%%%%%%%%%%%%%%%%%%%%%%%%%%%%%%%%%%%%%%%%%%%%%%%%%%%%%%%%%%%%%%%%%%%%%%%%%%%%%%%%%%%%%%%%%%%%%%%%%%%%%%%%%%%%%%%%%%%%%%%%%%

The most general four-pulse composite sequence has the form
\be\label{CP4}
(A_1)_{0} (A_2)_{\phi_2} (A_3)_{\phi_3} (A_4)_{\phi_4}.
\ee
We present three sets of four-pulse composite sequences, two symmetric and one asymmetric. 
%, following the pattern set by the three-pulse sequences.
%First, we consider the symmetric sequences and then the asymmetric ones.

%%%%%%%%%%%%%%%%%%%%%%%%%%%%%%%%%%%%%%%%%%%%%%%%%%%%%%%%%%%%
\subsubsection{Symmetric sequences of pulses}

The first symmetric sequence consists of identical nominal $\pi/2$ pulses (but with different phases)  \cite{Torosov2019},
\be\label{CP4a}
S4a:\quad (\tfrac12{\pi})_0 (\tfrac12{\pi})_{\frac12\pi} (\tfrac12{\pi})_{\frac32\pi-\theta} (\tfrac12{\pi})_{\pi-\theta},
\ee
where $\theta = 2\arcsin \sqrt{p}$. Its total pulse area is just $2\pi$.
The overall transition probability 
%$P = |U_{21}|^2$ 
reads
\be
P = p [1-\sin^4(\pi \epsilon/2)].
\ee
Obviously, it is accurate up to order $O(\eps^4)$. 
%This composite sequence has been derived in Ref.~\cite{Torosov2019}.

The other symmetric sequence of pulses reads \cite{Torosov2019}
\be\label{CP4b}
S4b:\quad (\tfrac12{\pi})_0 \pi_{\frac23\pi} \pi_{\frac53\pi-\theta} (\tfrac12{\pi})_{\pi-\theta}.
\ee
%where again $\theta = 2\arcsin \sqrt{p}$.
%
The overall transition probability reads
\be
P = p [1-\sin^6(\pi \epsilon/2)].
\ee
Obviously, in return to the larger total pulse area ot $3\pi$ compared to the previous sequence \eqref{CP4a} it is accurate up to the higher order $O(\eps^6)$. 
%If we set $P=\sin^2 (\theta/2)$ (as for a resonant $\theta$ pulse), we find $\psi = 2\chi -\theta$.

%Table \ref{Table:CP4} presents the explicit phases for the composite sequences \eqref{CP4a} and \eqref{CP4b} for a few values of the transition probability.
These sequences are very convenient as the availability of exact analytic formulae for the phases allows us to find their values for any value of the transition probability.

%F%F%F%F%F%F%F%F%F%F%F%F%F%F%F%F%F%F%F%F%F%F%F%F%F%F%F%F%F%F%F%F%F%F%F%F%F%F%F%F%F%F%F%F%F%F%F%F%F%F%F%F%F%F%F%F
\begin{figure}[tb]
\begin{tabular}{r}
    \includegraphics[width=0.90\columnwidth]{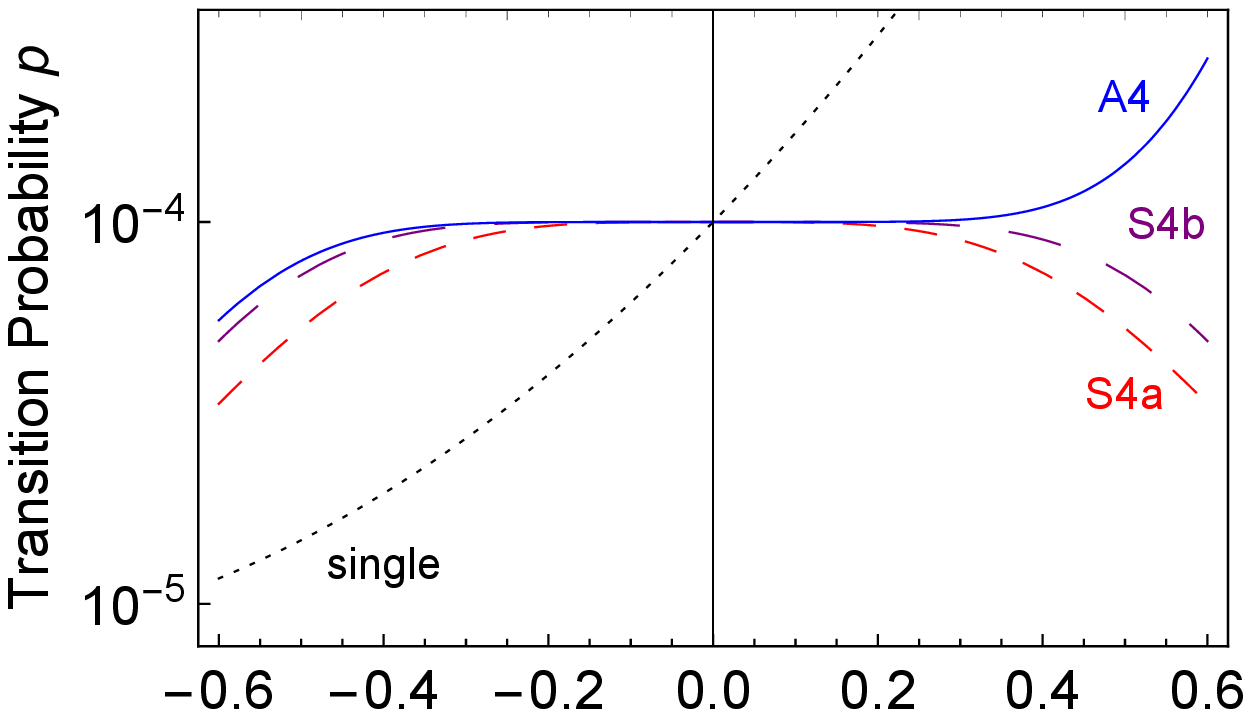} \\
    \includegraphics[width=0.90\columnwidth]{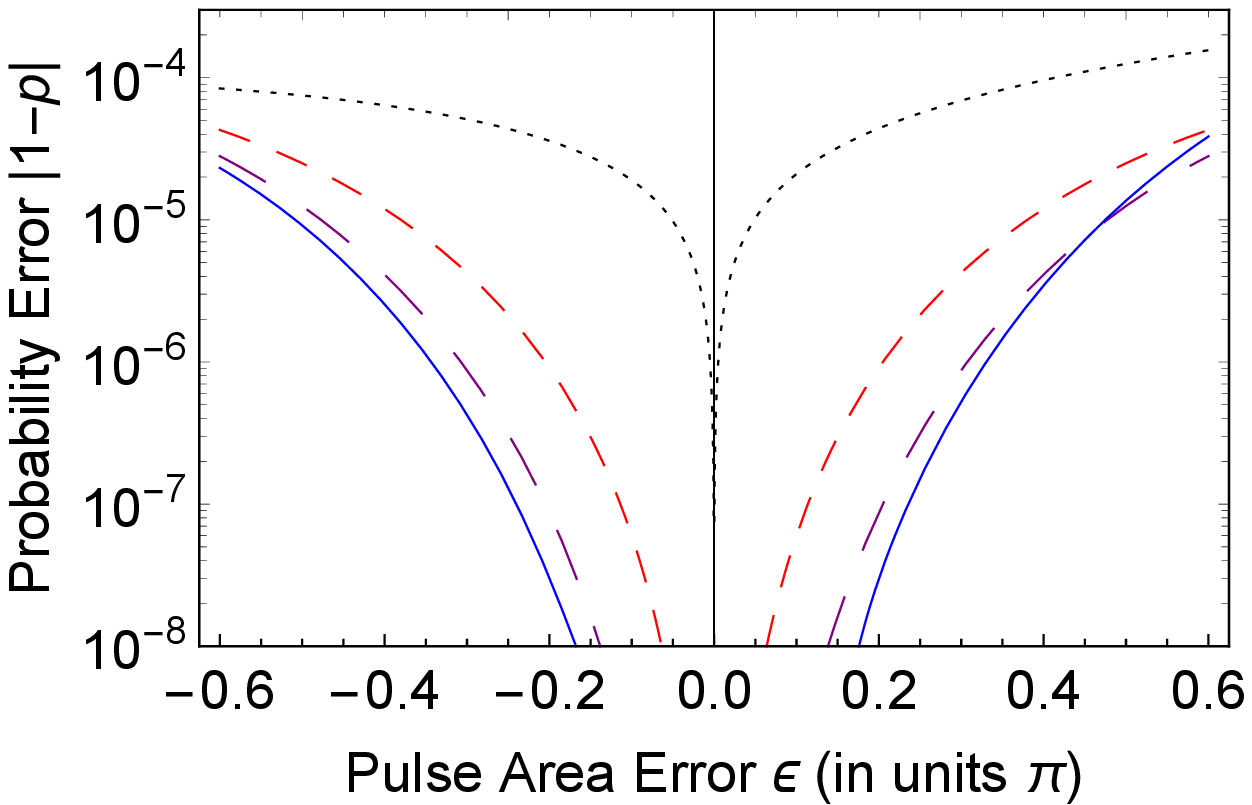}
\end{tabular}
\caption{
Performance of the four-pulse symmetrci composite sequences \eqref{CP4a} (red dashed), \eqref{CP4b} (purple long-dashed) and the asymmetric sequence \eqref{CP4c} (blue solid) for the transition probability $p=10^{-4}$.
The dotted curves show the single pulse excitation probability for comparison.
}
\label{Fig:cp4}
\end{figure}
%F%F%F%F%F%F%F%F%F%F%F%F%F%F%F%F%F%F%F%F%F%F%F%F%F%F%F%F%F%F%F%F%F%F%F%F%F%F%F%F%F%F%F%F%F%F%F%F%F%F%F%F%F%F%F%F

%%%%%%%%%%%%%%%%%%%%%%%%%%%%%%%%%%%%%%%%%%%%%%%%%%%%%%%%%%%%
\subsubsection{Asymmetric sequences}

The most general three-pulse composite sequence has the form
\be\label{CP4c}
%\alpha_0 \beta_\zeta \gamma_\upsilon.
A4:\quad (A_1)_{0} (A_2)_{\phi_2} (A_3)_{\phi_3} (A_4)_{\phi_4}.
\ee
All pulse areas and phases are free control parameters, which allow it to compensate a higher error order.
The pulse areas and the phases are computed numerically and are listed in Table \ref{Table:CP4}.
Although the asymmetric composite sequence \eqref{CP4c} costs more total pulse area ($\approx 3.44\pi$) than the preceding two sequences S4a and S4b, it is accurate to the higher order $O(\eps^7)$.

The performance of the four-pulse sequences is illustrated in Fig.~\ref{Fig:cp4}. 
All of them significantly outperform the single pulse profile and provide considerable stabilisation at the target transition probability value.
The best performance is delivered by the asymmetric sequence A4, which has the error order $O(\eps^7)$, followed by S4b, with the error order  $O(\eps^6)$, and then S4a, with the error order $O(\eps^4)$.
However, this ranking follows the total pulse area --- the cost factor --- which is $\approx 3.41\pi$ for A4, $3\pi$ for S4b, and $2\pi$ for S4a.
Note that the error order $O(\eps^4)$ for S4a is the same as the one for the three-pulse sequence $S3$ and one can verify that they generate similar excitation profiles.

%T%T%T%T%T%T%T%T%T%T%T%T%T%T%T%T%T%T%T%T%T%T%T%T%T%T%T%T%T%T%T%T%T%T%T%T%T%T%T%T%T%T%T%T%T%T%T%T
\begin{table}[t!]
\centering
%\footnotesize{
{
\begin{tabular}{|l|cccc|ccc|}
\hline
$p$ & $A_1$ & $A_2$ & $A_3$ & $A_4$ & $\phi_2$ & $\phi_3$ & $\phi_4$ \\
%\hline
%\multicolumn{9}{|c|}{Symmetric sequences of type S4a ($A_{\text{total}}=2\pi$)} \\ \hline 
%$10^{-2}$ & 4 & $\frac12$ & $\frac12$ & $\frac12$ & $\frac12$ & $\frac12$ & 1.2997 & 0.8997 \\
%$10^{-3}$ & 4 & $\frac12$ & $\frac12$ & $\frac12$ & $\frac12$ & $\frac12$ & 1.6685 & 0.5898 \\
%$10^{-4}$ & 4 & $\frac12$ & $\frac12$ & $\frac12$ & $\frac12$ & $\frac12$ & 1.4800 & 0.9800 \\
%$10^{-5}$ & 4 & $\frac12$ & $\frac12$ & $\frac12$ & $\frac12$ & $\frac12$ & 1.4937 & 0.9937 \\
%$10^{-6}$ & 4 & $\frac12$ & $\frac12$ & $\frac12$ & $\frac12$ & $\frac12$ & 1.4980 & 0.9980 \\
%$10^{-7}$ & 4 & $\frac12$ & $\frac12$ & $\frac12$ & $\frac12$ & $\frac12$ & 1.4994 & 0.9994 \\
%$10^{-8}$ & 4 & $\frac12$ & $\frac12$ & $\frac12$ & $\frac12$ & $\frac12$ & 1.4998 & 0.9998 \\
%\hline
%\multicolumn{9}{|c|}{Symmetric sequences of type S4b ($A_{\text{total}}=3\pi$)} \\ \hline 
%$10^{-2}$ & 6 & $\frac12$ & 1 & 1 & $\frac12$ & $\frac23$ & 1.6184 & 0.3821 \\
%$10^{-3}$ & 6 & $\frac12$ & 1 & 1 & $\frac12$ & $\frac23$ & 1.6685 & 0.5898 \\
%$10^{-4}$ & 6 & $\frac12$ & 1 & 1 & $\frac12$ & $\frac23$ & 1.7811 & 0.7234 \\
%$10^{-5}$ & 6 & $\frac12$ & 1 & 1 & $\frac12$ & $\frac23$ & 1.8538 & 0.8124 \\
%$10^{-6}$ & 6 & $\frac12$ & 1 & 1 & $\frac12$ & $\frac23$ & 1.9017 & 0.8725 \\
%$10^{-7}$ & 6 & $\frac12$ & 1 & 1 & $\frac12$ & $\frac23$ & 1.9337 & 0.9132 \\
%$10^{-8}$ & 6 & $\frac12$ & 1 & 1 & $\frac12$ & $\frac23$ & 1.9551 & 0.9409 \\
\hline
%\multicolumn{9}{|c|}{Asymmetric sequences of type W4 ($A_{\text{total}}=3\pi+\theta$)} \\ \hline
%$10^{-1}$ & 4 & 0.2048 & 1 & 1 & 1 & 0.9180 & 0.5980 & 1.4998 \\
%$10^{-2}$ & 4 & 0.0638 & 1 & 1 & 1 & 0.9538 & 0.7708 & 1.7216 \\
%$10^{-3}$ & 4 & 0.0201 & 1 & 1 & 1 & 0.9740 & 0.8701 & 1.8436 \\
%$10^{-4}$ & 4 & 0.0064 & 1 & 1 & 1 & 0.9853 & 0.9268 & 1.9120 \\
%$10^{-5}$ & 4 & 0.0020 & 1 & 1 & 1 & 0.9918 & 0.9588 & 1.9505 \\
%$10^{-6}$ & 4 & $6.4 \times 10^{-4}$ & 1 & 1 & 1 & 0.9954 & 0.9768 & 1.9722 \\
%$10^{-7}$ & 4 & $2.0 \times 10^{-4}$ & 1 & 1 & 1 & 0.9974 & 0.9870 & 1.9844 \\
%$10^{-8}$ & 4 & $6.4 \times 10^{-5}$ & 1 & 1 & 1 & 0.9985 & 0.9927 & 1.9912 \\
%\hline 
%\multicolumn{9}{|c|}{Asymmetric sequences of type A4 ($A_{\text{total}}=(2.8-3.0)\pi$)} \\ \hline
%$10^{-1}$ & 7 & 0.2048 & 1 & 1 & 1 & 0.9180 & 0.5980 & 1.4998 \\
$10^{-2}$ & 0.5367 & 1.1586 & 1.1360 & 0.5833 & 0.8499 & 1.5547 & 0.4360 \\
$10^{-3}$ & 0.8685 & 1.0434 & 0.3702 & 0.5174 & 1.0634 & 0.8847 & 0.0128 \\
$10^{-4}$ & 0.8165 & 0.9044 & 0.5579 & 0.6423 & 1.0362 & 0.9682 & 0.0146 \\
$10^{-5}$ & 0.7854 & 0.8335 & 0.6433 & 0.6905 & 1.0207 & 0.9856 & 0.0090 \\
$10^{-6}$ & 0.7669 & 0.7937 & 0.6875 & 0.7141 & 1.0118 & 0.9926 & 0.0052 \\
$10^{-7}$ & 0.7551 & 0.7698 & 0.7108 & 0.7255 & 0.9933 & 1.0042 & 1.9972 \\
$10^{-8}$ & 0.7494 & 0.7578 & 0.7244 & 0.7328 & 0.9962 & 1.0022 & 1.9984 \\
\hline
\end{tabular}
}
\caption{Pulse areas and phases (in units of $\pi$) for the composite sequences of 4 pulses \eqref{CP4c}.
%$(A_1)_{0} (A_2)_{\phi_2} (A_3)_{\phi_3} (A_4)_{\phi_4}$  for a few values of the transition probability $p$. 
%Also given is the order $n$ of pulse area error compensation $O(\epsilon^n)$, indicating that all orders below $n$ are zero.
All composite sequences have the error order $O(\epsilon^7)$.
 }
\label{Table:CP4}
\end{table}
%T%T%T%T%T%T%T%T%T%T%T%T%T%T%T%T%T%T%T%T%T%T%T%T%T%T%T%T%T%T%T%T%T%T%T%T%T%T%T%T%T%T%T%T%T%T%T%T

%%%%%%%%%%%%%%%%%%%%%%%%%%%%%%%%%%%%%%%%%%%%%%%%%%%%%%%%%%%%%%%%%%%%%%%%%%%%%%%%%%%%%%%%%%%%%%%%%%%%%%%%%%%%%%%%%%%%%%%%%%%%%%%%%%%%%%%%%
\subsection{Higher number of pulses}
%%%%%%%%%%%%%%%%%%%%%%%%%%%%%%%%%%%%%%%%%%%%%%%%%%%%%%%%%%%%%%%%%%%%%%%%%%%%%%%%%%%%%%%%%%%%%%%%%%%%%%%%%%%%%%%%%%%%%%%%%%%%%%%%%%%%%%%%%

Higher number of pulses present the opportunity for an error compensation of a higher order. 
There exist analytic symmetric composite sequences for arbitrary rotations, which can be used for small $p$ too \cite{Torosov2019}.
They are constructed as follows.
We can use a composite $\pi/2$ pulse to derive a composite $\theta$-pulse by applying a composite $\pi/2$ pulse sequence $C$,
 followed by the composite sequence  $C^R_{\theta}$, which is the time-reversed sequence $C$, with all its phases shifted by the same phase shift $\theta$,
\be\label{twin theta}
C_{0} C^R_{\theta},
\ee
an idea introduced by Levitt and Ernst \cite{Levitt1983}.
Moreover, if the sequence $C$ has the error order $O(\epsilon^n)$ then the composite $\theta$ sequence \eqref{twin theta} has the error order $O(\epsilon^{2n})$  \cite{Torosov2019}.
A few examples follow.

The composite sequence S2 of Eq.~\eqref{CP2a} becomes a composite $\pi/2$ pulse for $\theta=\pi/2$, which can be used in the twinning construction \eqref{twin theta},
\be\label{CP4d}
(\tfrac12\pi)_0 (\tfrac12\pi)_{\frac12\pi} (\tfrac12\pi)_{\frac32\pi-\theta} (\tfrac12\pi)_{\pi-\theta},
\ee
which is the same as the sequence S4a of Eq.~\eqref{CP4a}.
Because the sequence S2 has the error order $O(\epsilon^2)$ then the composite sequence S4a has the error order $O(\epsilon^4)$, as found in the previous section.

The composite sequence S3 of Eq.~\eqref{CP3a} for $\theta=\pi/4$ reads
\be
(\tfrac12{\pi})_0 \pi_{\frac34\pi} (\tfrac12{\pi})_{\pi},
\ee
and it has the error order $O(\epsilon^4)$.
By using the twinning construction \eqref{twin theta} we find a $\theta$ composite sequence of order $O(\epsilon^8)$,
\be\label{CP6s}
(\tfrac12{\pi})_0 \pi_{\frac34\pi} (\tfrac12{\pi})_{\pi}
(\tfrac12{\pi})_{2\pi-\theta} \pi_{\frac74\pi-\theta} (\tfrac12{\pi})_{\pi-\theta}.
\ee

One can build $\theta$ composite sequences of arbitrary length and arbitrary error order compensation by twinning the $\pi/2$ composite sequences \cite{Torosov2019}
\be\label{CP-symmetric}
(\pi/2)_{0} \pi_{\phi_2} \pi_{\phi_3} \cdots \pi_{\phi_{N-1}} (\pi/2)_{\phi_N},
\ee
composed of a sequence of $N-2$ nominal $\pi$ pulses, sandwiched by two pulses of areas $\pi/2$, with phases given by the analytic formula
\be
\phi_k = \frac{(k-1)^2}{2(N-1)}\pi \quad (k=1,2,\ldots,N).
\ee
It is easy to verify that the sequences \eqref{CP4d} and \eqref{CP6s} (after trivial population-preserving transformation of the phases) belong to such a family of sequences. 
Because the sequence \eqref{CP-symmetric} has the error order $O(\epsilon^{2(N-1)})$ the corresponding twinned sequence \eqref{twin theta} will have the error order $O(\epsilon^{4(N-1)})$.

Another, asymmetric family of  $\pi/2$ composite sequences can be used too \cite{Torosov2019},
\be\label{CP-asymmetric}
(\pi/2)_{0} \pi_{\phi_2} \pi_{\phi_3} \cdots \pi_{\phi_{N-1}} (\pi)_{\phi_N},
\ee
composed of a sequence of $N-1$ nominal $\pi$ pulses, preceded by a nominal $\pi/2$ pulse, with phases given by the analytic formula
\be
\phi_k = \frac{2(k-1)^2}{2N-1}\pi \quad (k=1,2,\ldots,N).
\ee
It has the error order $O(\epsilon^{2N-1})$.
Hence the twinning method \eqref{twin theta} generates $\theta$ sequences of the error order $O(\epsilon^{2(2N-1)})$.
For instance, for $N=3$ we find by twinning the $\theta$ sequence
\be\label{CP6a}
(\tfrac12{\pi})_0 \pi_{\tfrac25\pi} (\pi)_{\tfrac85{\pi}}
(\pi)_{\tfrac35\pi-\theta} \pi_{\tfrac75\pi-\theta} (\tfrac12{\pi})_{\pi-\theta},
\ee
which has the error order $O(\epsilon^{10})$.

Regarding the asymmetric composite sequences of 2, 3 and 4 pulses, presented above and derived numerically, it is computationally much harder to derive similar sequences for more than 4 pulses. Moreover, the advantage they deliver in terms of error order compensation for a given number of pulses compared to the symmetric sequences seems to decrease with the number of pulses $N$ and approach the point when the results do not repay the labour.

%%%%%%%%%%%%%%%%%%%%%%%%%%%%%%%%%%%%%%%%%%%%%%%%%%%%%%%%%%%%%%%%%%%%%%%%%%%%%%%%%%%%%%%%%%%%%%%%
%%%%%%%%%%%%%%%%%%%%%%%%%%%%%%%%%%%%%%%%%%%%%%%%%%%%%%%%%%%%%%%%%%%%%%%%%%%%%%%%%%%%%%%%%%%%%%%%
%%%%%%%%%%%%%%%%%%%%%%%%%%%%%%%%%%%%%%%%%%%%%%%%%%%%%%%%%%%%%%%%%%%%%%%%%%%%%%%%%%%%%%%%%%%%%%%%
%%%%%%%%%%%%%%%%%%%%%%%%%%%%%%%%%%%%%%%%%%%%%%%%%%%%%%%%%%%%%%%%%%%%%%%%%%%%%%%%%%%%%%%%%%%%%%%%
%%%%%%%%%%%%%%%%%%%%%%%%%%%%%%%%%%%%%%%%%%%%%%%%%%%%%%%%%%%%%%%%%%%%%%%%%%%%%%%%%%%%%%%%%%%%%%%%

\section{Quantum gates for ultrasmall rotations}

Ultrasmall rotation gates are more demanding to construct due to the necessity to have both the probabilities and the phases error-compensated. Mathematically, this is equivalent to expanding the propagator of the gate in a Taylor-Maclaurin series versus the error $\epsilon$ and set to zero the first few terms to the same error order $O(\epsilon^m)$ in all propagator matrix elements. Below we present several sequences, which produce high-fidelity rotation gates, two of which are known in the literature and one is derived here.

%%%%%%%%%%%%%%%%%%%%%%%%%%%%%%%%%%%%%%%%%%%%%%%%%%%%%%%%%%%%
\subsection{First-order error compensation}

The three-pulse rotation gate has been derived by Wimperis \cite{Wimperis1990},
\be\label{CP3w}
%\theta_0 \beta_{\phi} \beta_{3\phi},
W3:\quad \theta_0 \pi_{\phi} \pi_{3\phi},
\ee
%with $\beta = \pi$,
with $\theta = \arccos (1-2p) = 2\arcsin\sqrt{p}$ and $\phi = \arccos(-\theta/(2\pi)) \approx \frac12 \pi + \sqrt{p}$.
It is accurate up to order $O(\eps^2)$.
It is a phase-distortionless sequence and hence suitable for a rotation gate.

Another three-pulse rotation gate has the form \cite{Gevorgyan2021}
\be\label{G3}
G3:\quad \alpha_{\phi_1} \pi_{\phi_2} \alpha_{\phi_1},
\ee
where $\alpha$ is determined from the equation
\be
\frac{\pi\sin (\alpha)}{\alpha} = 2\cos(\theta/2).
\ee
Given $\alpha$, we can find $\phi_1$ and $\phi_2$ from 
\bse
\begin{align}
%&-\sin (\alpha) \cos (\phi _1-\phi _2) = \cos(\theta/2), \label{R3-1} \\
2 \alpha \cos (\phi _1-\phi _2) + \pi &= 0, \label{R3-3} \\
 \sin (\phi _1-\phi _2) &= \sin (\theta/2) \cos (\phi _1).
\end{align}
\ese
This composite sequence is related to the SCROFULOUS composite pulse \cite{Cummins2003} and it is accurate to the error order $O(\eps^2)$.

The values of the pulse area and the composite phases are given in Table \ref{Table:rotation-3}.

%T%T%T%T%T%T%T%T%T%T%T%T%T%T%T%T%T%T%T%T%T%T%T%T%T%T%T%T%T%T%T%T%T%T%T%T%T%T%T%T%T%T%T%T%T
\begin{table}
\centering
\begin{tabular}{|c|lll|}
\hline
 \multicolumn{4}{c} Rotation gate G3: $(\frac12\pi+x)_{\phi_1} \pi_{\pi+y} (\frac12\pi+x)_{\phi_1}$ \\
\hline
 $p$ & $x$ & $\phi_1$ & $y$ \\
\hline
$10^{-2}$ & $2.5\times 10^{-3}$ & $2.492\times 10^{-2}$ & $5.672\times 10^{-2}$ \\
$10^{-3}$ & $2.5\times 10^{-4}$ & $7.904\times 10^{-3}$ & $1.797\times 10^{-2}$ \\
$10^{-4}$ & $2.5\times 10^{-5}$ & $2.500\times 10^{-3}$ & $5.683\times 10^{-3}$ \\
$10^{-5}$ & $2.5\times 10^{-6}$ & $7.906\times 10^{-4}$ & $1.797\times 10^{-3}$ \\
$10^{-6}$ & $2.5\times 10^{-7}$ & $2.500\times 10^{-4}$ & $5.684\times 10^{-4}$ \\
\hline
\end{tabular}
\caption{
Parameters of the composite sequence G3 of Eq.~\eqref{G3} for different transition probabilities $p$.
}
\label{Table:rotation-3}
\end{table}
%T%T%T%T%T%T%T%T%T%T%T%T%T%T%T%T%T%T%T%T%T%T%T%T%T%T%T%T%T%T%T%T%T%T%T%T%T%T%T%T%T%T%T%T%T

%%%%%%%%%%%%%%%%%%%%%%%%%%%%%%%%%%%%%%%%%%%%%%%%%%%%%%%%%%%%
\subsection{Second-order error compensation}

A well-known composite seqeunce, which compensates the second-order error is the BB1 sequence of Wimperis \cite{Wimperis1994},
\be\label{BB1}
\text{BB1} = (\pi/2)_{0} \pi_{\chi} (2\pi)_{3\chi} \pi_{\chi},
\ee
with $\chi=\arccos(-\theta/4\pi)$.
It produces arbitrary phase-distortionless rotations at the angle $\theta$ with the error order $O(\epsilon^3)$

%%%%%%%%%%%%%%%%%%%%%%%%%%%%%%%%%%%%%%%%%%%%%%%%%%%%%%%%%%%%%%%%%%%%%%%%%%%%%%%%%%%%%%%%%%%%%%%%%%%%%%%%%%%%%%%%%%%%%%%%%%%%%%%%%%%%%%%%%%%%%%%%%%%%%%%%%%%%%%%%

%%%%%%%%%%%%%%%%%%%%%%%%%%%%%%%%%%%%%%%%%%%%%%%%%%%%%%%%%%%%%%%%%%%%%%%%%%%%%%%%%%%%%%%%%%%%%%%%%%%%%%%%%%%%%%%%%%%%%%%%%%%%%%%%%%%%%%%%%
\section{Conclusions}
%%%%%%%%%%%%%%%%%%%%%%%%%%%%%%%%%%%%%%%%%%%%%%%%%%%%%%%%%%%%%%%%%%%%%%%%%%%%%%%%%%%%%%%%%%%%%%%%%%%%%%%%%%%%%%%%%%%%%%%%%%%%%%%%%%%%%%%%%

We presented a solution to the problem of generating well-defined very small excitation of a two-state quantum transition.
The method uses composite pulse sequences of two, three, four and more pulses. 
Both symmetric and asymmetric, analytic and numeric classes of sequences have been presented and analyzed in detail. 

%For two- and three-pulse sequences, the solutions are analytic and applicable to any transition probability.
%For more than three pulses we have only been able to find numerical solutions.

The results in this paper can be useful in application such as single-photon generation by a cold atomic ensemble of $N$ atoms. A composite sequence producing a transition probability of $1/N$ will make sure that only one excitation is shared within the ensemble, to be subsequently released by a scheme like DLCZ. 
Another possible application is fine tuning of quantum gates, in which accurate small adjustments of the rotation angle are needed in order to reach high fidelity. 
Yet another application is the generation of huge Dicke states in cold atomic ensembles or trapped ions by global collective addressing.

\acknowledgments
HG acknowledges support from the EU Horizon-2020 ITN project LIMQUET (Contract No. 765075).
NVV acknowledges support from  the Bulgarian national plan for recovery and resilience, contract BG-RRP-2.004-0008-C01 (SUMMIT), project number 3.1.4.

%%%%%%%%%%%%%%%%%%%%%%%%%%%%%%%%%%%%%%%%%%%%%%%%%%%%%%%%%%%%%%%%%%%%%%%%%%%%%%%%%%%%%%%%%%%%%%%%%%%%%%%%%%%%%%%%%%%%%%%%%%%%%%%%%%%%%%%%%%%%%%%%%%%%%%%%%%%%%%%%
%%%%%%%%%%%%%%%%%%%%%%%%%%%%%%%%%%%%%%%%%%%%%%%%%%%%%%%%%%%%%%%%%%%%%%%%%%%%%%%%%%%%%%%%%%%%%%%%%%%%%%%%%%%%%%%%%%%%%%%%%%%%%%%%%%%%%%%%%%%%%%%%%%%%%%%%%%%%%%%%
%%%%%%%%%%%%%%%%%%%%%%%%%%%%%%%%%%%%%%%%%%%%%%%%%%%%%%%%%%%%%%%%%%%%%%%%%%%%%%%%%%%%%%%%%%%%%%%%%%%%%%%%%%%%%%%%%%%%%%%%%%%%%%%%%%%%%%%%%%%%%%%%%%%%%%%%%%%%%%%%

\end{document}